# Rawls' Fairness, Income Distribution and Alarming Level of Gini Coefficient†

Yong Tao[a], Xiangjun Wu[b], Changshuai Li[c]

[a]College of Economics and Management, Southwest University, China
[b]School of Economics, Huazhong University of Science and Technology, China
[c]Jonn and Walk Department of Economics, Clemson University, USA

**Abstract**: The argument that the alarming level of Gini coefficient is 0.4 is very popular, especially in the media industry, all around the world for a long time. Although the 0.4 standard is widely accepted, the derivation of the value lacks rigid theoretical foundations. In fact, to the best of our knowledge, it is not based on any prevalent and convincing economic theories. In this paper, we incorporate Rawls' principle of fair equality of opportunity into Arrow-Debreu's framework of general equilibrium theory with heterogeneous agents, and derive the alarming level of Gini coefficient formally. Our theory reveals that the exponential distribution of income not only satisfies Pareto optimality, but also obeys social fairness in Rawls' sense. Therefore, we specify the maximal value of the Gini coefficient when income follows exponential distribution as a possible alarming level. Our computations show that the alarming level should be specified at least equal or larger than 0.5 rather than 0.4. We empirically investigate if our model receives support from a large data set of all kinds of countries all over the world from Word Bank in 1990, 1995, 2000 and 2005 using the distribution fitting and statistical decision methodology. The results suggest that the value of 0.4 is around the mean of the Gini coefficients, corresponding to the most probable event in a peaceful world, rather than the alarming level, while the two-sigma rule shows that in our sample the alarming levels are all larger than 0.5, conforming to the predictions of our theory.

---

†Project supported by the Scholarship Award for Excellent Doctoral Student Granted by Ministry of Education of China (2012), (Grant No. 0903005109081-019); Natural Science Foundation of China (Grant No. 71271096); National Excellent Doctor Fund (Grant No.201304).

# 1. Introduction

Over the last half century, since the seminal work of Kuznets (1955), there has emerged a vast and well-development literature on the theory and empirical studies of income inequality, though the controversy is still pervasive. Empirical support for the existence of substantial income inequality and their importance in generating social, political and economic instabilities has made the introduction of such variables into models of macroeconomics the prevalent and accepted way of modeling social instability. Given the broad interest in studying the effects of income inequality on macroeconomic aggregates and its determinants, economists have been constructing and testing all kinds of theoretical models and empirical studies. However, the questions in the theoretical models and empirics are far from settled. The popular approach now is to use the share of total income accruing to some parts of the top income holders, such as the top 10% group, to measure the income concentration (see, e.g., Piketty and Saez, 2014). Some other papers are more prone to use some indexes, such as the Gini coefficient, to gauge the overall inequality; see Anand and Segal (2008) for an excellent discussing of the methodological issues on the measurement of global inequality. A comparison between these two strands of measures, see Alvaredo (2011). Unfortunately, these papers do not discuss what types of income distributions will unite efficiency and fairness, provided that the latter ensures social stability. In this paper, we introduce Rawls' fairness considerations into general equilibrium model to study its implications for the statistical distribution of income for a competitive economy. Then, based on the income distribution, we try to derive the secure scope of the Gini coefficient.

The motivation for introducing Rawls' fairness considerations comes from experimental, empirical studies etc, which provide a wealth of evidence supporting the assumption or idea of the fair equality of opportunity (Bolton et al., 2005). Although some evidence indicates that allocation fairness and procedural fairness are linked in important ways, they are conceptually distinct. A considerable amount of work has been directed at allocation fairness (Kagel and Roth, 1995), and the critical finding is that fair outcomes are often enforced by resistance to unfair outcomes. There is also substantial evidence showing that how individuals view inequality and redistribution is heavily influenced by the perceived source of inequality, see, e.g., Benabou and Tirole (2006), and Durante et al. (2014). Ku and Salmon (2013) show that individuals' willingness to accept higher but more unequal outcomes depends on the source of the initial inequality and random assignment leads to the most tolerance for disadvantageous inequality. However, considering the theme in this paper we focused on, we are trying to seek the income distributions with the highest probability from the long-run competitive process, rather than analyze the ultimate competitive outcomes. So, the focal point in this paper is centered on the procedural fairness.

Our baseline model is a standard general equilibrium model with consumers and firms. In that framework, due to the free entry assumption there are multiple equilibria in the economy. To that framework, we incorporate Rawls' fairness, as an additional consideration, into the model to seek the income distribution with the highest probability from the long-run competitive equilibrium allocations. Once we get the statistical distribution of the income, we can derive the corresponding Gini coefficient. There are many ways to express and calculate the Gini coefficient from the individual data (see, e.g., Dorfman (1979); Atkinson and Bourguignon (2000); Xu (2000, 2007); Barrett and Donald (2009)). However, the most of the calculations depend on the income distribution, if the income distribution of the total population is not available, the estimation will not be precise. Unlike the previous mentioned papers, in this paper, we will derive the statistical distribution of the income under rigid economic theory.

The Gini coefficient is commonly used as a measure of inequality of income distribution between a nation's residents. It is defined as a ratio with values between 0 and 1, the larger the value, the higher the inequality and vice versa. A Gini coefficient of zero corresponds to perfect income equality, where everyone has an exactly the same income; meanwhile, one corresponds to perfect income inequality, i.e. one person has all the income, while everyone else has zero income. Since the income inequality is often regarded as the cause of social instability, the Gini coefficient is naturally identified as the early-warning signal.

For a long time, the international institutions, such as the World Bank, UN, and the news media, etc, accept that the alarming level of Gini coefficient should be set at 0.4 (UNRISD, 2013). Some authors also implicitly hint that 0.4 is a critical value of the Gini coefficient for the more developed countries (MDCs) (Biancotti, 2006). This means that if the Gini coefficient of one country exceeds 0.4, it may confront the risk of overall social instability. However, in the economics profession, economists even cannot make clear where this alarming level come from! Indeed, such an alarming level is not developed on the basis of any available and convincing economic theories, let alone on rigid empirical tests.

It is well known that, Arrow-Debreu's general equilibrium model can be regarded as the standard tool for dealing with the optimal resource allocation among social members. The basic proposition is: Competitive equilibrium corresponds to the equilibrium income allocation. By the first fundamental theorem of welfare economics, the competitive equilibrium is Pareto optimal, so is the equilibrium income allocation. Nevertheless, "Pareto optimality" does not necessarily imply the existence of "fairness". A Pareto optimal allocation may be very unfair (Pazner and Schmeidler, 1974; Alesina et al., 2005; Alesina et al., 2012). Due to this disappointing result, some welfare economists suggest that searching the best equilibrium income allocation through a so-called social welfare function that reflects collective preferences in a society. Unfortunately, Arrow's Impossibility Theorem refuses the existence of such a social welfare function (Arrow, 1963). This is just the well-known "dilemma of social choice".

Although Arrow's Impossibility Theorem casts a shadow on the availability of democratic decision, Rawls (1999) opened another fascinating outlet to deal with the problem of social fairness. To guarantee social fairness, Rawls (1999) introduces the principle of equal liberty and the principle of fair equality of opportunity. The first principle governs the assignment of rights and duties, while the second governs the distribution of income and wealth. In this paper, we attempt to incorporate Rawls' principle of fair equality of opportunity into Arrow-Debreu's general equilibrium model. Our purpose is to seek an income distribution which not only satisfies Pareto optimality but also obeys Rawls' principle of fair equality of opportunity. Since such an income distribution ensures both efficiency and fairness, we can specify the maximal value of the corresponding Gini coefficient as a possible alarming level.

To get some intuition, we simply present the basic idea of our method here. Assuming that a competitive economy produces four equilibrium income allocations $\{A_1, A_2, A_3, A_4\}$, each of them is Pareto optimal. Meanwhile, we further assume that these four equilibrium allocations can be divided into the following three income distributions: $a_1 = \{A_1\}$, $a_2 = \{A_2, A_3\}$ and $a_3 = \{A_4\}$. By Rawls' principle of fair equality of opportunity, we know that each equilibrium allocation occurs with an equal probability 1/4 (Tao 2013), then we conclude that $a_1$ occurs with the probability $1/4$, $a_2$ occurs with the probability $1/2$, and $a_3$ occurs with the probability $1/4$. This means that the income distribution $a_2$ occurs with the highest probability, so it can be regarded as a "certain event" in a just society[1]. It is easy to show that income distribution $a_2$ not only satisfies Pareto optimality but also obeys Rawls' principle of fair equality of opportunity: Because $A_2$ and $A_3$ are Pareto optimal, $a_2 = \{A_2, A_3\}$ is also Pareto optimal. The probability of the occurrence of $a_2$ is the largest can be resorted to the Rawls' principle of fair equality of opportunity.

To sum up, one can use three steps to seek the income distribution with the highest probability (Tao 2013). First, try to find all possible equilibrium income allocations of a competitive economy. Second, divide all these equilibrium income allocations into different income distributions. Finally, find the income distribution which contains the most equilibrium income allocations, which occurs with the highest probability. The main purpose of this paper is to seek the income distribution with the highest probability in a long-run competitive economy using the three steps stated above, and then specify the maximal value of the Gini coefficient of this income distribution as a possible alarming level.

The paper is organized as follows: Section 2 defines the long-run competitive equilibrium under the framework of Arrow-Debreu economy, and proves that the long-run competitive economy will produce multiple equilibrium income allocations. Section 3 introduces the Rawls' principle of fair equality of opportunity to the model, and divides the equilibrium income allocations into different income distributions. Section 4 shows that the income distribution with the highest probability in the long-run competitive economy follows an exponential distribution, and meanwhile, we find that it leads to a possible alarming level of Gini coefficient specified at least equal or larger than 0.5. Section 5 describes our data set, empirical methods, and presents some evidence that support the theoretical predictions of our model. Finally, section 6 sums up the results and concludes.

## 2. The model

The basic model we propose to understand the income distribution combines the behavior of firms and consumers under the general equilibrium framework. The quite standard microeconomic tools are discussed in detail in Tao (2013). The novel features of our model are that the introduction of a framework for exploring the income distributions between heterogeneous agents, and incorporating fairness into the derivation of income distribution. We first show the existence of long-run competitive equilibria and later reformulate them in the form of income allocations.

### 2.1 Assumptions

Following the framework of neoclassical economics[2] (Mas-Collel et al., 1995; Page 579), we assume that there are $N$ consumers, $N$ firms and $L$ types of commodities in the economy.

#### 2.1.1 Consumers

The basic assumptions of the consumer behavior are as follows:

---

[1] If the number of the equilibrium allocations is large enough, then the "highest probability" tends to 1. Then the income distribution with the highest probability is indeed a "certain event" in a just society.
[2] Our model is undoubtedly a special case of Arrow-Debreu's general equilibrium model, since we have assumed that the number of firms equals the number of consumers.

(a) Each consumer $i = 1, ..., N$ faces with some possible consumption bundles in some set, the consumption set $X_i \subset R^L$, $X_i$ is closed, convex, and bounded below. There is no satiation consumption bundle for any consumer. We denote the consumption vector of the $i$th consumer by $x_i = (x_{1i}, ..., x_{Li})$, where $x_i \in X_i$ and $x_{ki} \geq 0$ for $k = 1, ..., L$.

(b) A preference relation $\succsim_i$ defined on $X_i$. For each consumer, the sets $\{x_i \in X_i | x_i \succsim_i x'_i\}$ and $\{x_i \in X_i | x'_i \succsim_i x_i\}$ are closed.

(c) If $x_i^1$ and $x_i^2$ are two arbitrary points in $X_i$, and $t \in (0,1)$, then $x_i^2 \succsim_i x_i^1$ implies that $tx_i^2 + (1-t)x_i^1 \succ_i x_i^1$.

(d) There is $x_i^0$ in $X_i$, such that $x_i^0 \ll \omega_i$, where the $\omega_i \in R^L$ is an initial endowment vector for the consumer $i$.

### 2.1.2 Firms

The basic assumptions of the firms are as follows:

(e) Each firm $j = 1, ..., N$ is endowed with a production set $Y_j \subset R^L$. We denote the production vector of the $j$th firm by $y_j = (y_{1j}, ..., y_{Lj})$, where $y_j \in Y_j$.

(f) $0 \in Y_j$.

(g) $Y = \sum_{j=1}^{N} Y_j$ is closed and convex.

(h) $-Y \cap Y = \{0\}$.

(i) $-R_+^L \subset Y$.

Since our purpose is to use Arrow-Debreu's general equilibrium model (ADGEM, hereafter) to deal with the income allocation problem between consumers, we need to make the following two extra assumptions:

(j). Without loss of generality, we assume that all the firms only produce the $m$th type of commodity, namely $y_{mj} \geq 0$ for $j = 1, ..., N$ and $y_{lj} \leq 0$ for $l \neq m$. The implication of such an assumption is that there is only one industry in the economy.

(k). We assume that the $i$th consumer is the owner of the $i$th firm, so the revenue of the $i$th firm is the income of the $i$th consumer, where $i = 1, ..., N$.

Besides the assumptions (a) to (i), the production technology satisfies additivity and public available technology, that is, $Y_j + Y_j \subset Y_j$, for any $j$, and $Y_1 = Y_2 = ... Y_N$. The implications of additivity and public available technology see assumption 3.1 and assumption 3.2 in Tao (2013). Tao (2013) has shown that ADGEM with additivity and publicly available technology has multiple equilibria, and can be identified with the long-run competitive economy. From now on we will always call ADGEM with additivity and publicly available technology the long-run competitive economy. In accordance with Piketty and Saez (2014), who study the long-run evolutionary process of the income inequality, we also consider the long-run features of income distribution in this paper. So, in this regard, it is the long-run competitive economy that we considered.

## 2.2 Long-run competitive equilibria

In the spirit of Marshall, long-run competition is identified with free entry. Because free entry implies that equilibrium profit of any firm is zero (Varian, 1992, 2003), Tao (2013) proposes the following definition of the long-run competitive equilibria:

***Definition 1***: An allocation $(x_1^*, ..., x_N^*; y_1^*, ..., y_N^*)$ and a price vector $P = (P_1, ..., P_L)$ constitute a long-run competitive equilibrium if the following three conditions are satisfied:

(1). Profit maximization: For each firm $i$, there exists $y_i^* \in Y_i$ such that $P \cdot y_i \leq P \cdot y_i^* = 0$

for all $y_i \in Y_i$.

(2). Utility maximization: For each consumer $i$, $x_i^* \in X_i$ is the solution of maximizing the preference $\succsim_i$ under the budget set: $\{x_i \in X_i : P \cdot x_i \leq P \cdot \omega_i\}$.

(3). Market clearing: $\sum_{i=1}^{N} x_i^* = \sum_{i=1}^{N} \omega_i + \sum_{i=1}^{N} y_i^*$.

We now give the following crucial propositions.

***Proposition 1***: If the assumptions (a) to (i) are satisfied, long run competitive economy has multiple equilibria:
$$\left(x_1^*, \ldots, x_N^*; y_1^*(t_1), \ldots, y_N^*(t_N)\right), \tag{1}$$
where $y_i^*(t_i) = t_i z^*$ for $i = 1, \ldots, N$, meanwhile $x_i^*$ for $i = 1, \ldots, N$ and $z^* = (z_1^*, \ldots, z_L^*)$ are fixed vectors.

Moreover, $\{t_i\}_{i=1}^{N}$ satisfies:
$$t_i \geq 0, \text{ for } i = 1, \ldots, N, \text{ subject to } \sum_{i=1}^{N} t_i = 1. \tag{2}$$
The $z^*$ is called the total production vector and obeys the equality:
$$P \cdot z^* = 0. \tag{3}$$
**Proof**. By Proposition 3.3 in Tao (2013) completes this proof. □

***Proposition 2***: Each of the multiple equilibrium states in equation (1) is Pareto optimal.

**Proof**. By first fundamental theorem of welfare economics we immediately get the result. □

**2.3 Equilibrium income allocations**

Equation (1) and (3) together illustrate that each firm $i$ in the long-run equilibria only obtains zero economic profit. Then by assumption (j) and $y_i^*(t_i) = t_i z^*$ for $i = 1, \ldots, N$, the $i$th firm will obtain $t_i P_m z_m^*$ units of revenue, where $z_m^*$ denotes the $m$th component of $z^*$. By assumption (k) the consumer $i$ is the owner of the firm $i$, so the consumer $i$ will obtain $t_i P_m z_m^*$ units of income. Therefore, the equilibrium income allocation among $N$ consumers can be written in the form:
$$(t_1 P_m z_m^*, \ldots, t_N P_m z_m^*). \tag{4}$$
If we denote the total income by $\Pi = P_m z_m^*$, the equilibrium income allocations $(t_1 P_m z_m^*, \ldots, t_N P_m z_m^*)$ can be directly written as:
$$(R_1, R_2, \ldots, R_N), \tag{5}$$
where $R_i$ denotes the income of the $i$th consumer and by equation (2) it satisfies:
$$R_i \geq 0, \text{ for } i = 1, \ldots, N, \text{ with } \sum_{i=1}^{N} R_i = \Pi. \tag{6}$$
Undoubtedly, by Proposition 2, we know that any equilibrium income allocation satisfying equation (5) and (6) is Pareto optimal. In fact, equation (5) and (6) will be the starting point of the following discussions.

**3. Fairness axiom and income distribution**

In section 2, we have shown that, equation (5) and (6), the long-run competitive economy will produce multiple equilibrium allocations, and proposition 2 indicates that each of the equilibrium allocations is Pareto optimal. However, "Pareto optimality" does not imply "fairness". For example, the equilibrium income allocation $(\Pi, 0, \ldots, 0)$ satisfies equation (6), but it also indicates maximal inequality (one consumer holds all income $\Pi$). To avoid such a dilemma, some welfare economists propose searching the best equilibrium income allocation outcome by making use of a so-called social welfare function, which aggregates individual preferences. Unfortunately,

Arrow's Impossibility Theorem refused the existence of the social welfare function (Jehle and Reny, 2001; Page 243).

**3.1 Fairness axiom**

Although Arrow's Impossibility Theorem casted a shadow on the availability of democratic decision, Rawls (1999) opened another fascinating outlet to deal with the problem of social fairness. Rawls argued that a just economy can be regarded as a fair procedure which will translate its fairness to the (equilibrium) outcomes, so that every social member is indifferent between these outcomes. With this idea, Rawls (1999; Page 76) introduced the principle of fair equality of opportunity. This principle indicates that each equilibrium outcome of a fair economy should be selected with equal opportunities as collective decisions[3], to go one step further, Rawls' principle of fair equality of opportunity can be expressed as the following fairness axiom (Tao 2013):

*Axiom 1*: If a competitive economy produces $\omega$ equilibrium income allocations, and at the same time, the economy is absolutely fair, then each equilibrium income allocation occurs with an equal probability $1/\omega$.

Now we show that, by the Fairness Axiom 1, an income distribution with the highest probability not only satisfies Pareto optimality but also obeys Rawls' principle of fair equality of opportunity. To this end, we consider a simple case where a long-run competitive economy produces a set of equilibrium allocations $\{A_1, A_2, A_3, A_4\}$, each of them is Pareto optimal. We further assume that these four equilibrium allocations can be divided into the following three kinds of income distributions[4]: $a_1 = \{A_1\}$, $a_2 = \{A_2, A_3\}$ and $a_3 = \{A_4\}$. Then by the fairness axiom 1, we know that each equilibrium allocation occurs with the probability $1/4$, so we conclude that $a_1$ also occurs with the probability $1/4$, by the same token, $a_2$ occurs with the probability $1/2$, while $a_3$ occurs with the probability $1/4$. Clearly, the income distribution $a_2$ occurs with the highest probability, so it can be regarded as the "certain event" in a just society[5]. It is easy to show that $a_2$ satisfies Pareto optimality and obeys Rawls' principle of fair equality of opportunity[6].

**3.2 Income distribution**

In this section, we will concentrate our attention on the long-run competitive economy. The discussion proposed in section 3.1 implies that by the Fairness Axiom 1 we can use three steps to seek the income distribution with the highest probability. First, try to find all possible equilibrium income allocations of a competitive economy. Second, divide all these equilibrium income allocations into different income distributions. Finally, find the income distribution containing the most equilibrium income allocations, which occurs with the highest probability.

To apply the Fairness Axiom 1 into the long-run competitive economy, we must divide the multiple equilibrium allocations (5) into different income distributions. By the method proposed in Appendix A, the multiple equilibrium allocations (5) can be divided into different income distributions $\{a_k\}_{k=1}^n$, which follows the following definition:

---

[3] By (5) and (6) this means that every social member will have an equal chance of occupying any given income level.

[4] $a_1 = \{A_1\}$ represents an income distribution when equilibrium allocation $A_1$ occurs. Likewise, $a_2 = \{A_2, A_3\}$ and $a_3 = \{A_4\}$ are defined in the same way. Concrete example sees the Example 1 in Appendix A.

[5] Here we only consider a simple case with four equilibrium allocations. In fact, if the number of the equilibrium allocations is large enough, then the "highest probability" will tend to 1.

[6] Because $A_2$ and $A_3$ are Pareto optimal, $a_2 = \{A_2, A_3\}$ is of course Pareto optimal as well. $a_2$ occurs with the largest probability dues to the Fairness Axiom 1, so it obeys Rawls' principle of fair equality of opportunity.

***Definition 2***: We denote the set of all possible equilibrium income allocations satisfying equation (5) and (6) by $W$. We call the non-negative number sequence, $\{a_k\}_{k=1}^n = \{a_1, a_2, \ldots, a_n\}$, an income distribution if and only if it is a subset of $W$, as well as obeys the following four conventions:

(1). There are totally $n$ possible income levels in the economy: $\varepsilon_1 < \varepsilon_2 < \cdots < \varepsilon_n$;

(2). There are $a_k$ consumers, each of them obtains $\varepsilon_k$ units of income, where $k$ ranges from 1 to $n$;

(3). $\sum_{k=1}^n a_k = N$;

(4). $\sum_{k=1}^n a_k \varepsilon_k = \Pi$.

For a given income distribution $\{a_k\}_{k=1}^n$, we immediately know that, there are $a_1$ consumers, each of them obtains $\varepsilon_1$ units of income, while for the $a_2$ consumers, each of them obtains $\varepsilon_2$ units of income, …, and so on.

The Appendix A further shows that a given income distribution $\{a_k\}_{k=1}^n$ contains $\Omega(\{a_k\}_{k=1}^n)$ equilibrium income allocations, where $\Omega(\{a_k\}_{k=1}^n)$ is denoted by:

$$\begin{cases} \Omega(\{a_k\}_{k=1}^n) = \frac{N!}{\prod_{k=1}^n a_k!} \\ \sum_{k=1}^n a_k = N \\ \sum_{k=1}^n a_k \varepsilon_k = \Pi \end{cases} \quad (7)$$

Therefore, by the third step proposed at the beginning of this subsection, seeking the income distribution with the highest probability, $\{a_k^*\}_{k=1}^n$, is equivalent to solving the following maximization problem (Tao 2013):

$$\begin{cases} \underset{\{a_k\}_{k=1}^n}{Max}: \Omega(\{a_k\}_{k=1}^n) \\ s.t. \sum_{k=1}^n a_k = N \\ \sum_{k=1}^n a_k \varepsilon_k = \Pi \end{cases} \quad (8)$$

## 4. The results

By the Lemma 6.2 in Tao (2013), the maximization problem (8) is equivalent to the following problem:

$$\begin{cases} \underset{\{a_k\}_{k=1}^n}{Max}: ln\Omega(\{a_k\}_{k=1}^n) \\ s.t. \sum_{k=1}^n a_k = N \\ \sum_{k=1}^n a_k \varepsilon_k = \Pi \end{cases} \quad (9)$$

Substituting (7) into (9) we obtain the income distribution with the highest probability (in fact, denoted by the probability which tends to one[7]):

$$a_k = \frac{1}{e^{\alpha + \beta \varepsilon_k}} \quad (10)$$

where $k = 1, 2, \ldots, n$, and $\alpha \leq 0$, $\beta \geq 0$ (see Tao (2010, 2013)). The complete solutions will be given in Appendix B.

The exponential distribution (10) is also called the Boltzmann-Gibbs distribution. It is worth mentioning that due to the Fairness Axiom 1, equation (10) arises because the society is assumed to be absolutely fair. However, the human society cannot be absolutely fair, so the exponential distribution (10) may be only suitable for a part of the population. Indeed, Yakovenko et al. (2009) use the income data from U.S. in 1983–2000, and confirm that the income of the majority of population (lower class) obey the exponential distribution (or Boltzmann-Gibbs distribution), see Figure 1. Not coincidently, Nirei et al. (2007) also find the same empirical result by using the income data from Japan, see Figure 2.

---

[7] It is easy to check that when $N \to \infty$ by the formula (7) the "highest probability" (Tao 2013), $P[\{a_k^*\}_{k=1}^n] = \frac{\Omega(\{a_k^*\}_{k=1}^n)}{\sum_{\{a_k'\}_{k=1}^n} \Omega(\{a_k'\}_{k=1}^n)}$, will tend to 1, where we denote by $P[X]$ the probability that an event $X$ occurs. By Tao's spontaneous order theory, the exponential distribution (10) is obviously a spontaneous order in the sense of collective behaviors about income allocations (Tao 2013).

Moreover, from Figures 1 and 2 we further notice that the income of a small fraction of population (upper class) in the society obeys Pareto distribution (or power distribution). It is well known that, Pareto distribution can be derived using some unfair rules, e.g., the rule of "The rich get richer" (Barabasi, 1999). By the same method proposed by Barabasi (1999), we can easily get Pareto's income distribution (Tao 2014):

$$a_k = \varepsilon_k^{-\gamma-1}, \tag{11}$$

where $k = 1,2,...,n$, and $\gamma \geq 1$.

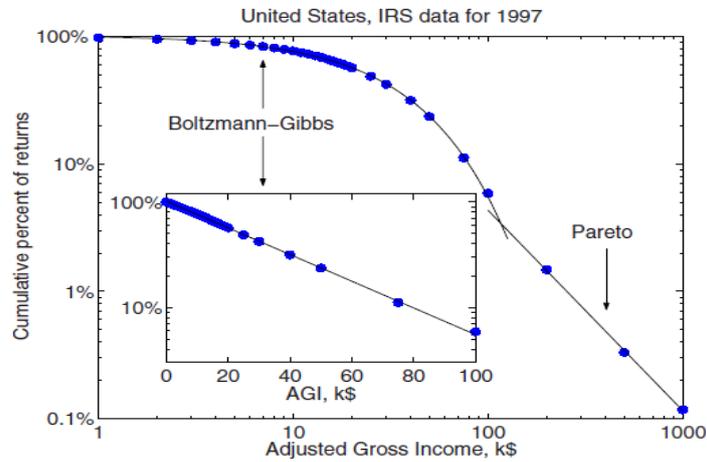

**Figure 1**: Reprinted from Yakovenko et al. (2009). Points represent the Internal Revenue Service data, and solid lines are fits to Boltzmann-Gibbs and Pareto distributions.

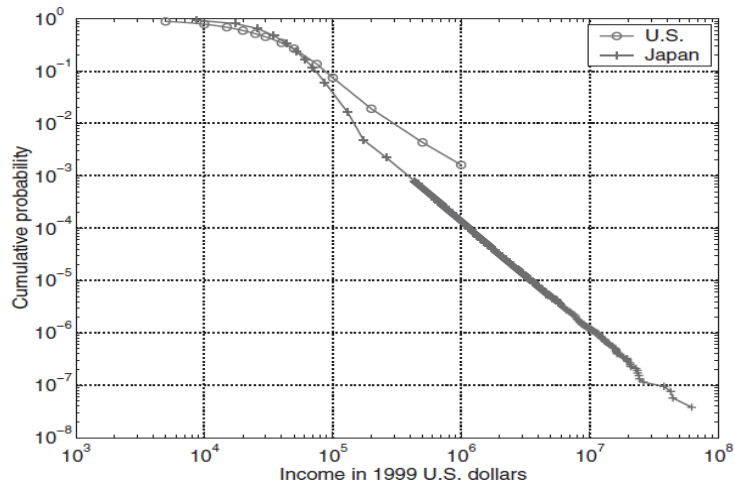

**Figure 2**: Reprinted from Nirei et al. (2007). Income distributions in the U.S. and Japan in 1999.

## 4.2 The alarming level of Gini coefficient

On the one hand, by the discussion of section 3.1, we know that the exponential distribution (10) can be regarded as a signal of indicating social fairness, since it is a result of Rawls' principle of fair equality of opportunity.

On the other hand, the exponential distribution obeys Pareto optimality as well, since by Definition 2, it is a set of equilibrium income allocations, each of which is Pareto optimal.

To sum up, the exponential distribution (10) not only ensures efficiency but also guarantees fairness, and therefore can be regarded as a signal of indicating social stability. Due to this, we can specify the maximal value of the Gini coefficient of the exponential distribution (10) as the potential alarming level.

Next we attempt to calculate the Gini coefficients of the exponential distribution (10) and Pareto distribution (11) respectively. To this end, let us rewrite (10) and (11) in the form of continuous functions:

$$f_B(\varepsilon) = \begin{cases} \beta e^{-\beta\left(\varepsilon+\frac{\alpha}{\beta}\right)}, & \varepsilon \geq -\frac{\alpha}{\beta} \\ 0, & \varepsilon < -\frac{\alpha}{\beta} \end{cases} \tag{12}$$

$$f_P(\varepsilon) = \begin{cases} \gamma a^\gamma \varepsilon^{-\gamma-1}, & \varepsilon \geq a \\ 0, & \varepsilon < a \end{cases}, \tag{13}$$

where, $f_B(\varepsilon)$ and $f_P(\varepsilon)$ correspond to exponential distribution (10) and Pareto distribution (11) respectively, and $\varepsilon$ represents the continuous variable describing the possible income level.

By the techniques proposed in Appendix C, we can calculate the Gini coefficients of the exponential distribution $f_B(\varepsilon)$ and the Pareto distribution $f_P(\varepsilon)$ separately as follows:

$$G_B = \frac{1}{2(1-\alpha)}, \tag{14}$$

$$G_P = \frac{1}{2\gamma-1}, \tag{15}$$

We know that $\alpha \leq 0$, $\gamma \geq 1$, so the intervals of Gini coefficients $G_B$ and $G_P$ are:

$$0 \leq G_B \leq 0.5, \tag{16}$$

$$0 \leq G_P \leq 1, \tag{17}$$

By expression (16) we surprisingly find that the exponential distribution (10) rules out the extreme inequality, for example, the equilibrium income allocation $(\Pi, 0, \dots, 0)$ (whose Gini coefficient equals 1) is ruled out. However, by expression (17) we note that the Pareto distribution (11) cannot rule out the allocation $(\Pi, 0, \dots, 0)$. This is because the exponential distribution (10) is a result based on the Fairness Axiom 1 but the Pareto distribution (11) is a result based on the "The rich get richer" rule which may involve unfair behaviors.

Since the exponential distribution (10) not only ensures efficiency[8] but also guarantees fairness, it can be regarded as a powerful signal of indicating social stability in the viewpoint of normative economics. This means, the income follows the exponential distribution implies that the society is stable. As we have shown in expression (16), when income follows the exponential distribution, and the world is free of extreme inequality, the scope of the corresponding Gini coefficient lies in the interval between 0 and 0.5, so once the Gini coefficient lies in this interval, the society is stable in a large way. Whenever the value goes out of this scope, that is, it is larger than the maximal value, instability may occur.

We have known that the derivation of the Pareto distribution (11) is based on the unfair and non-equilibrium rules (i.e., some rules like "The rich get richer"), so we cannot guarantee that a society whose income follows power distribution will be fair at least in Rawls' sense. That is to say, we cannot guarantee that every member in such a society will have an equal chance of occupying any given income level. Although the Pareto distribution (11) may ensure efficiency, it contradicts social fairness at least in Rawls' sense. As a result, under Rawls' fairness perspective, the alarming level of Gini coefficient should be only considered when income follows exponential distribution.

So we specify the maximal value of Gini coefficient $G_B$ when income follows exponential distribution, which equals 0.5, as a minimal basic reference point of the alarming level, clearly violating the international standard 0.4. We will give the empirical evidence in the next section.

## 5. Some empirical evidence

In this section, we report some empirical results that support the theoretical predictions of our model. To test the implications of the model we use a sub-sample of the World Bank's PovcalNet database that allows us to analyze some aspects of the alarming level of Gini coefficient and to understand the role played by the alarming on economic uncertainty.

### 5.1. Data description and methodology

Before presenting the data and methods of our empirical studies, we need to make the following assumption and proposition, which are the foundations of our proceeding investigations.

*Assumption 1*: In times of peace, the political instability around the world is a small probability event.

*Proposition 3*: If there are no political or economic interventions among countries, the distribution of Gini coefficients follows the normal distribution.

**Proof**. See Appendix D. □

Assumption 1 just states that in times of peace, only very small number of countries may experience the political instability. We do not rule out the possibility of the emergence of political instability, but just think that the instability is not a system event. Put in a statistical way, if the sample size is big enough, the observations undergo political instability are only some negligible outliers. So we can fit some stable distributions on the data and use the statistical decision theory to make the inference.

---

[8] By Definition 2 we note that the exponential distribution (10) consists of the equilibrium income allocations, and hence lies in the core of the economy, another strand of literature from game theory also proves the akin argument (see, e.g., Howe and Roemer, 1981)

Proposition 3 implies that the Gini coefficient follows stable normal distribution when the sample size is big enough, and large adverse shocks are very rare. In fact, only when the samples follow some stable distributions can we make credible statistical inferences, or conclusions based on unstable distributions are not reliable.

The sample in our study includes data from more than 130 countries all over the world that cover different stages of economic development and time spans. However, due to the availability of data, we can only collect the relative fully data sets from four separate years, that is: 1990, 1995, 2000 and 2005. In 1990, we get 130 observations, while in 1995, it is 137, and in 2000, 2005, the observations are 139 and 140 separately. All the data are in the form of percentage.

The only variable in our empirical study is the Gini coefficient, which are collected from Word Bank Database. In order to test our hypothesis, we first resort to the Jarque-Bera Chi-square statistic to test the normality of the data, and then under the stable empirical distribution, we use the statistical decision theory to detect the alarming level.

The test statistic for normality of observations was proposed by Jarque and Bera (1980, 1987). The statistic can be written as follows:

$$JB = \frac{n}{6}\left(S^2 + \frac{1}{4}(K-3)^2\right), \qquad (18)$$

where $n$ is the number of observations (or degrees of freedom in general); $S$ is the sample skewness, and $K$ is the sample kurtosis. If the data comes from a normal distribution, the $JB$ statistic asymptotically has a chi-squared distribution with two degrees of freedom, so the statistic can be used to test the hypothesis that the data are from a normal distribution. The null hypothesis is a joint hypothesis of the skewness being zero and the excess kurtosis being zero. Samples from a normal distribution have an expected skewness of zero and an expected excess kurtosis of zero (which is the same as a kurtosis of 3). As the definition of $JB$ shows, any deviation from this increases the $JB$ statistic. For small samples the chi-squared approximation is overly sensitive, often rejecting the null hypothesis when it is in fact true. Furthermore, the distribution of p-values departs from a uniform distribution and becomes a right-skewed uni-modal distribution, especially for small p-values. This leads to a large Type I error rate. So, to circumvent the small sample problem, we desert the data sets whose observations are small, and choose some representative years when the sample size is relatively large.

In regarding to the alarming level of the Gini coefficient, we follow the approach of classical statistical decision theory, that is, the so-called three-sigma rule (Bartoszynsk and Niewiadomska-Buga, 2008). According to the rule, we are allowed to disregard the possibility of a random variable deviating from its mean more than three standard deviations. Namely, if $X \sim N(\mu, \sigma^2)$, where $\mu$ is the mean, $\sigma^2$ is the variance, then

$$P\{|X - \mu| > 3\sigma\} = P\left\{\left|\frac{X-\mu}{\sigma}\right| > 3\right\} = 0.0026. \qquad (19)$$

From equation (19), we know that for any random variable sampled from normal distribution, the probability that it will deviate from its mean by more than three standard deviations is about 3‰, which is, at most, very small and negligible. Consider that the symmetric feature of the normal distribution, and what we care in this paper is the right tail, the probability becomes even smaller, 1.5‰. so this rule is helpful for us to drop some outliers in the sample, while at the same time without losing much information, but in this paper, our purpose is to detect the alarming level of Gini coefficient, rather than making observation selection, so we make some compromises on the rule and use a variant–the two-sigma rule, that is, we treat the random variables deviating from their means more than two standard deviations as the small probability events, meaning that the occurrence of the events are possible, though not probable. Following the same formula as equation (19), we can easily find that the probability of a random variable deviating from its mean more than two standard deviations is 0.0455, in the right tail, it is 0.0228. Any events happening at this or smaller probabilities, we regard them as the small probability events. Based on this argument, we can calculate the critical value of the alarming of Gini coefficient empirically. The happening of small probability events implies that some countries are undergoing unstable systems, showing that economic or political crisis may emerge.

**5.2 Empirical results: A tale of two worlds**

In order to verify and examine the predictions of our theory, we follow the methods of distribution fitting and normality test just proposed, the statistical tests are reported in Figure 3–6. Each figure depicts the results of this exercise for the four separate years.

We start by considering the empirical distribution of the data. As the Figure 3–6 shows, which reports the Jarque-Bera chi-squared statistic and the corresponding p-values for all the cases studied. Next we turn to the alarming level of the Gini coefficient. The calculation method is just the two-sigma rule. Just like the three-sigma rule, we use the following formula:

$$P\{|X - \mu| > 2\sigma\} = P\left\{\left|\frac{X-\mu}{\sigma}\right| > 2\right\} = 0.0455,$$

So $X = \mu + 2\sigma$ is just the critical value, we only consider the right tail of the distribution, so when the real data is larger than this value, the small probability event happens, showing that the country or society begins to go to unstable.

**5.2.1 World in 1990**

In the year of 1990, the summary statistics of Gini coefficient presented in Figure 3 show that the null hypothesis of normality of the distribution can be rejected at the usual 5% marginal significance level. This result is very interesting and insightful, while at the same time, consistent with our theory prediction (see proposition 3). The history of 1990 across the world was in fact very dark and full of uncertainty, politics and economy went into chaos, a series of astonishing incidents occurred, such as the reunification of Germany, the Gulf war and the Baltic states declaring independence from the Soviet Union, et al., to name a few. When all these events reflected in the data, the result is that the Gini coefficient no long follows a stable normal distribution. Due to this, we have no way to calculate the alarming level, so we do not report the value in this year. Although the mean of the Gini coefficient in 1990 is smaller than that of the other years, the standard deviation is much larger, a signal of unstable.

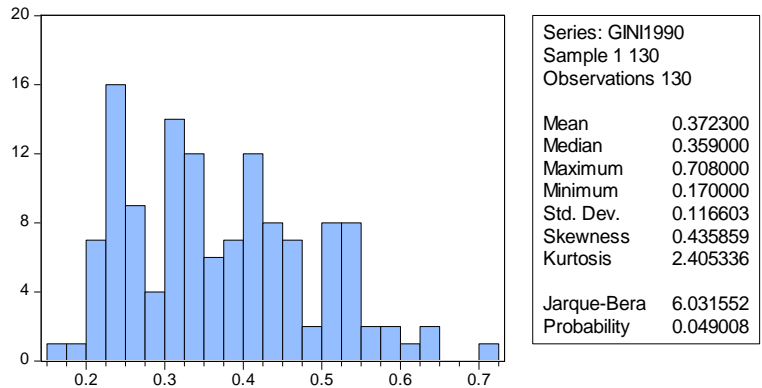

**Figure 3**: The histogram and summary statistics of Gini coefficient in 1990

### 5.2.2 World in 1995, 2000, and 2005

When it comes to the year of 1995, 2000, and 2005, it is very clear that the null hypothesis of normality of the distribution cannot be rejected at the usual 5% marginal significance level, in fact, the P-value reported in these three years are all larger than 10%, a strong signal that the null hypothesis should not be rejected. Rather than 1990, the world economy and politics in 1995, 2000, and 2005 were in the states of euphoria.

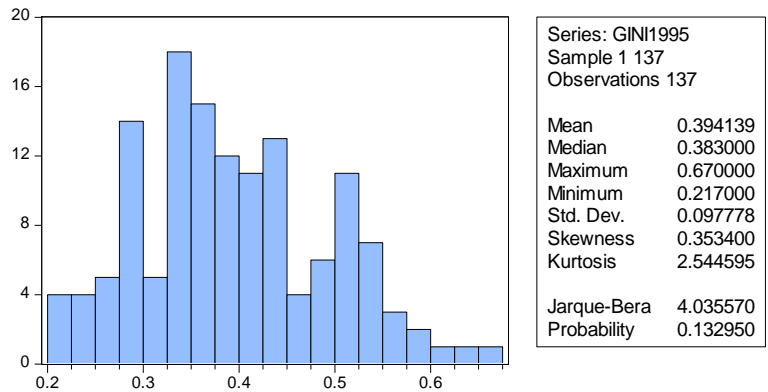

**Figure 4**: The histogram and summary statistics of Gini coefficient in 1995

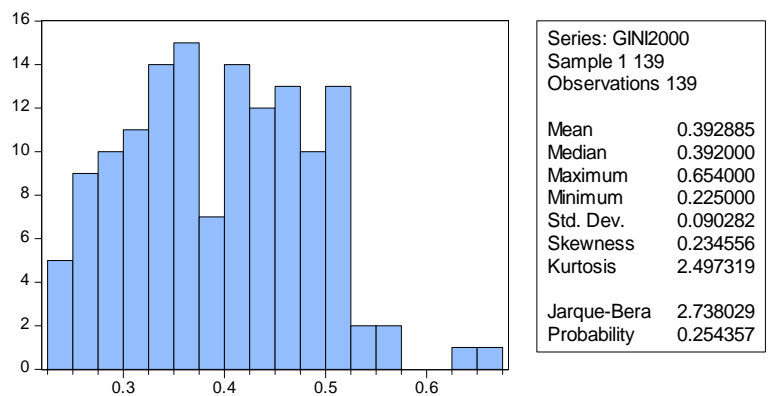

**Figure 5**: The histogram and summary statistics of Gini coefficient in 2000

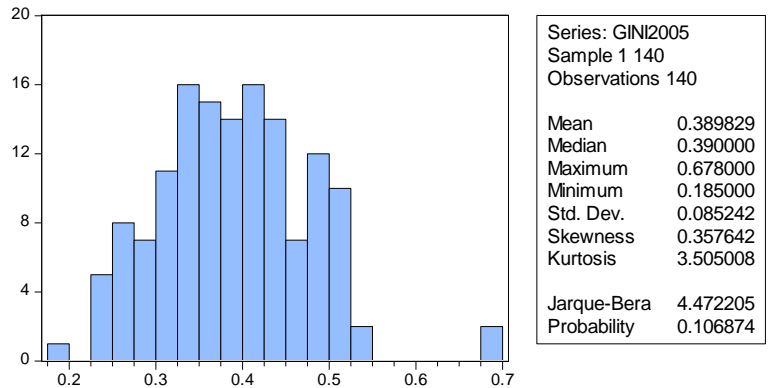

**Figure 6**: The histogram and summary statistics of Gini coefficient in 2005

The normality distribution tests in Figure 4–6 implies stable distributions exist in 1995, 2000, and 2005. So we can calculate the alarming level of Gini coefficients for the three years. Resorting to a little computation, the values are 0.579695, 0.573449, and 0.560313，corresponding separately to the year of 1995, 2000, and 2005. The results are summarized in Table 1.

Table 1: The alarming levels of Gini coefficient

| Year | Alarming level of Gini coefficient |
|---|---|
| 1995 | 0.579695 |
| 2000 | 0.573449 |
| 2005 | 0.560313 |

Note: The data of Gini coefficient in 1990 no longer follows normal distribution at the 5% significance level, so the alarming level is not reported.

Although the choice of data (they are spanned by 5 years) makes it difficult to establish a link between the economy and politics uncertainty, it seems possible to find a relationship between the above-stated coefficients and the fluctuations of economic activity. As one can recognize from the calculation, the critical values of Gini coefficients, inferred from the numerical fitting, tend to decrease in our sample periods, the degree of stability around the world in fact decreases.

The distribution and the critical values of the Gini coefficient we presented in this section are clearly compatible with our theory. From the Figures 3–6 above, one can see that in normal years, the distributions of the data are very stable, while in the turbulent year, the stable distribution no longer exists. As for the alarming level, in contrast to the international standard, no one coefficient lies around 0.4, in fact, they are all much larger. Our results show that the traditional measuring of the alarming level is intrinsically fault, in our sample, 0.4 is just the mean of the Gini coefficients, implies that it is the most probable value, rather than the critical value, so the traditional standard cannot fully reflect the facts of income or wealth inequality.

In summary, the proposition that the alarming level of Gini coefficient at least equal or larger than 0.5 is in agreement with our empirical observations, thus, when Gini coefficient goes up, but not yet cross the alarming level, the government should pay more attention to the economic development, rather than inequality alleviation.

## 6. Concluding remarks

We set out in this paper to shed light on Rawl's fairness, income distribution and the alarming level of Gini coefficient both in a theoretical and empirical perspective. Due to the lacking of rigid theory on the traditional argument of the alarming level of the Gini coefficient, we develop a theoretical model that captures the essence of Rawls' fairness and income distribution under the general equilibrium framework. Based on the model, we get the theoretical critical value of alarming level of Gini coefficient.

In the theoretical model, the income distribution we derived from the general equilibrium model not only satisfies Rawls' fairness but also stays in the core of the economy, and the distribution follows the exponential distribution exactly. To put it in another way, the exponential distribution satisfies the Pareto optimality, while at the same time, conforms to Rawls' fairness, it is the direct result of the fair competition in the society. However, when it comes to the reality, the problem arises: human society can never be absolutely fair, so we conjecture that in countries with mature and sound legal systems and democratic regimes, the majority of the population, their income follows the exponential distribution; while for the minority of the people, their income level follows non-fair distribution, according to the usual "the rich get richer" rule, which is obviously non-fair, we know that the distribution is the Pareto distribution.

In fact, our theoretical predictions are supported by empirical facts. Studies on the income distribution of the U.S. and Japan show that in both countries the income of the majority of the people follows the exponential distribution, while the income of the minority follows the Pareto distribution, meaning that the income distribution in a society is of two forms. So our theoretical assertions are confirmed.

Based on the theoretical predictions obtained from our model, we test the theory using the data of the Gini coefficient collected from all kinds of countries all over the world in four separate years. We first show that the distributions of Gini coefficient all over the world are normal distributed in the states of euphoria, while in the turbulent year, it is no longer following a stable normal distribution. This result presents an implication for seeking the potential alarming level of Gini coefficient when regional conflict is a small probability event in the peaceful world, but when the world undergo radical changes or turbulences, the level cannot be calculated from a stable distribution. Next we calculate the alarming levels of Gini coefficients from three years data, that is, 1995, 2000, and 2005, using the statistical decision theory. The results suggest that the alarming levels are not only much larger than 0.4, but also larger 0.5, supporting the proposal posed in our theoretical model. An interesting exception is that in 1990, the Gini coefficient no longer follows the normal distribution at the 5% significance level, though not appropriate, we still informally calculate the alarming level under normal distribution using the two sigma rule, the value is 0.605506, implying that when a country's Gini coefficient is larger than 0.6, it is going to become more unstable, just like the year of 1990.

# Appendices

**Appendix A**

In this appendix, we attempt to divide the multiple equilibrium income allocations, that is, the equation (5), into different income distributions. For the sake of simplicity, we drop the constraint $\sum_{i=1}^{N} R_i = \Pi$ temporarily.

Let's first consider a simple case with only two heterogeneous consumers:

***Example 1***: Two consumers and two possible income levels: $\varepsilon_1 < \varepsilon_2$.

We first explore how many equilibrium income allocations will example 1 have. Due to the fact that there are only two consumers, we need to count all the possible income allocations $(R_1, R_2)$ satisfying equation (6) for $N = 2$. However, because the constraint $\sum_{i=1}^{2} R_i = \Pi$ has assumed to be temporarily dropped, we only need to count all possible income allocations $(R_1, R_2)$ satisfying $R_i \geq 0$ for $i = 1,2$. As a result, there are four equilibrium income allocations:

$A_1 = (\varepsilon_2, \varepsilon_2)$, $A_2 = (\varepsilon_1, \varepsilon_2)$, $A_3 = (\varepsilon_2, \varepsilon_1)$, $A_4 = (\varepsilon_1, \varepsilon_1)$.

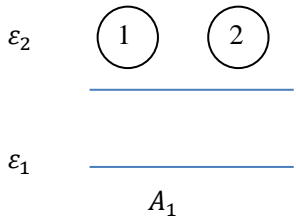

**Figure 7**: Equilibrium income allocation $A_1$: consumers 1 and 2 obtains $\varepsilon_2$ units of income respectively.

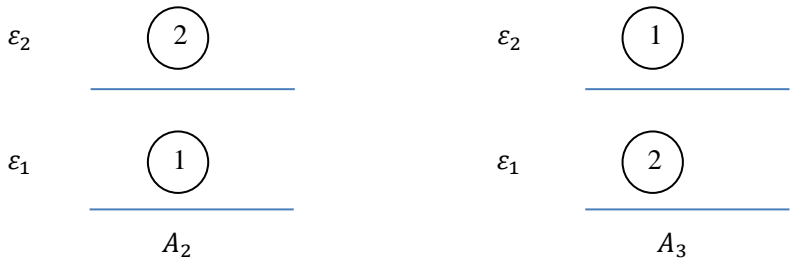

**Figure 8**: Equilibrium income allocation $A_2$: consumer 1 obtains $\varepsilon_1$ units of income; consumer 2 obtains $\varepsilon_2$ units of income. Equilibrium income allocation $A_3$: consumer 1 obtains $\varepsilon_2$ units of income; consumer 2 obtains $\varepsilon_1$ units of income.

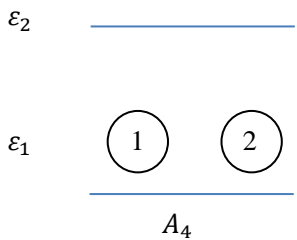

**Figure 9:** Equilibrium income allocation $A_4$: consumers 1 and 2 obtain $\varepsilon_1$ units of income respectively.

If we denote a consumer by a ball, then each equilibrium income allocation $A_i (i = 1,2,3,4)$ can be shown by a figure, see Figure 7–9. For example, the Figure 7 depicts the equilibrium income allocation $A_1$ in which consumers 1 and 2 obtain $\varepsilon_2$ units of income respectively, where the ball 1 stands for consumer 1 and ball 2 stands for consumer 2.

Now we know that the Example 1 has four possible equilibrium income allocations: $A_1$, $A_2$, $A_3$ and $A_4$; each of which can be shown by a figure. If one observes the Figures 7–9 carefully, then one may find that these four equilibrium allocations can be divided into three different groups. To see this, we consider an ordered pair $\{a_1, a_2\}$, where $a_1$ represents that there are $a_1$ consumers each of who obtains $\varepsilon_1$ units of income, and $a_2$ represents that there are $a_2$ consumers each of who obtain $\varepsilon_2$ units of income. Adopting this notion, one easily finds that the Figures 7–9 can be just denoted by $\{a_1 = 0, a_2 = 2\}$, $\{a_1 = 1, a_2 = 1\}$ and $\{a_1 = 2, a_2 = 0\}$ respectively.

It is easy to see that $A_2$ and $A_3$ will obey a unified rule: One consumer obtains $\varepsilon_1$ units of income and the other obtains $\varepsilon_2$ units of income. Therefore, both $A_2$ and $A_3$ satisfy the ordered pair $\{a_1 = 1, a_2 = 1\}$. This means that an ordered pair $\{a_1, a_2\}$ can be thought of as a 'set' whose elements are equilibrium income allocations. For example, $A_2$ and $A_3$ obey the ordered pair $\{a_1 = 1, a_2 = 1\}$, so we get:

$$\{a_1 = 1, a_2 = 1\} = \{A_2, A_3\}. \tag{A.1}$$

Similarly, we have:

$$\{a_1 = 0, a_2 = 2\} = \{A_1\}. \tag{A.2}$$

$$\{a_1 = 2, a_2 = 0\} = \{A_4\}. \tag{A.3}$$

If we extend the analysis of two consumers to $N$ consumers, then we have the definition about income distribution; that is, the Definition 2 in sub-section 3.2.

We denote the number of elements in a given income distribution $\{a_k\}_{k=1}^n$ by $\Omega(\{a_k\}_{k=1}^n)$. For the case of heterogeneous consumers, by Boltzmann statistics method (Tao; 2010, 2013), we immediately know that the number of elements in a given income distribution $\{a_k\}_{k=1}^n$ is denoted by:

$$\Omega(\{a_k\}_{k=1}^n) = \frac{N!}{\prod_{k=1}^n a_k!}. \tag{A.4}$$

Example 1 indicates that $N = 2$ and $n = 2$. Thus, using the formula (A.4) we can compute the number of elements in each income distribution as follows:

$$\Omega(\{a_1 = 0, a_2 = 2\}) = \frac{2!}{0! \times 2!} = 1, \tag{A.5}$$

$$\Omega(\{a_1 = 1, a_2 = 1\}) = \frac{2!}{1! \times 1!} = 2, \tag{A.6}$$

$$\Omega(\{a_1 = 2, a_2 = 0\}) = \frac{2!}{2! \times 0!} = 1, \tag{A.7}$$

Clearly, the results (A.5)-(A.7) are consistent with the numbers of equilibrium income allocations listed by Figures 7–9 respectively.

Let us now recover the constraint $\sum_{i=1}^N R_i = \Pi$, so the correct expression of $\Omega(\{a_k\}_{k=1}^n)$ can be expressed as follows:

$$\begin{cases} \Omega(\{a_k\}_{k=1}^n) = \frac{N!}{\prod_{k=1}^n a_k!} \\ \sum_{k=1}^n a_k = N \\ \sum_{k=1}^n a_k \varepsilon_k = \Pi \end{cases} \tag{A.8}$$

**Appendix B**

In this appendix, we will derive the solutions of equation (9).

By the formula (A.4), the function $ln\Omega(\{a_k\}_{k=1}^n)$ can be written in the form:

$$ln\Omega(\{a_k\}_{k=1}^n) = lnN! - \sum_{k=1}^n lna_k!. \tag{B.1}$$

Due to the fact that the value of $a_k$ is large enough, using the Stirling's formula (Carter, 2001; Page 218):

$$lnm! = m(lnm - 1), \quad (m \gg 1), \tag{B.2}$$

the equation (B.1) can be rewritten in the form:

$$ln\Omega(\{a_k\}_{k=1}^n) = N(lnN - 1) - \sum_{k=1}^n a_k(lna_k - 1). \tag{B.3}$$

The method of Lagrange multiplier for the optimal problem (9) gives

$$\frac{\partial ln\Omega(\{a_k\}_{k=1}^n)}{\partial a_k} - \alpha \frac{\partial N}{\partial a_k} - \beta \frac{\partial \Pi}{\partial a_k} = 0, \quad k = 1,2,\ldots,n \tag{B.4}$$

Substituting (B.3) into (B.4) we obtain the income distribution with the highest probability:

$$a_k = \frac{1}{e^{\alpha+\beta\varepsilon_k}}, \quad k = 1,2,\ldots,n. \tag{B.5}$$

where $\alpha \leq 0$ and $\beta \geq 0$ (Tao 2010, 2013).

**Appendix C**

In this appendix, we will derive the expression of the Gini coefficient under exponential and Pareto distribution respectively, that is, the equation (14) and (15).

We assume that the income level $x$ in an arbitrary country is a continuous variable, and lies in a closed interval $[a,b]$, where $a \geq 0$ and $b < +\infty$, the probability density function (PDF) is $f(x)$, while the cumulative distribution function (CDF) is $F(x)$, according to the definitions of probability distribution, $F(x)$ is the percentage of population whose income less than $x$, that is:

$$F(x) = \int_a^x f(x)\, dt. \tag{C.1}$$

Then the mean of the income $\mu$ is:

$$\mu = \int_a^b xf(x)\, dx. \tag{C.2}$$

Under the settings assumed above, the formula for calculating the much-used Gini coefficient $G$ can be written as (see, for example, Lambert, 1993, p. 43):

$$G = \frac{2}{\mu} \int_a^b x\left[F(x) - \frac{1}{2}\right] f(x)dx. \tag{C.3}$$

Now we use the formula (C.3) to calculate the Gini coefficients of exponential distribution and power distribution respectively.

By equation (12) the exponential distribution is as follow:

$$f_B(x) = \begin{cases} \beta e^{-\beta\left(x+\frac{\alpha}{\beta}\right)}, & x \geq a, \\ 0, & x < a \end{cases} \tag{C.4}$$

where $\alpha \leq 0, \beta \geq 0$ and $a = -\frac{\alpha}{\beta}$.

Substituting (C.4) into (C.3) and by order $b \to \infty$ we obtain:

$$G_B = \frac{1}{2(1-\alpha)}. \tag{C.5}$$

By (13) the power distribution is as follow:

$$f(x) = \begin{cases} \gamma a^{\gamma} x^{-\gamma-1}, & \gamma \geq a \\ 0, & \gamma < a \end{cases}, \qquad (C.6)$$

where $\gamma \geq 1$.

Substituting (C.6) into (C.3) and by order $b \to \infty$ we obtain:

$$G_P = \frac{1}{2\gamma - 1}. \qquad (C.7)$$

## Appendix D

In this appendix, we complete the proof of the Proposition 3 in the manner by which Gauss (1809) shows that the measurement error obeys normal distribution.

In a market-oriented economy, it is naturally supposed that by the "invisible hand" the economy would tend to a desirable Gini coefficient. We can denote the desirable Gini coefficient by $G_D$. However, there would be many and many different factors which force the actual Gini coefficient to deviate from the desirable Gini coefficient. For example, there are different resource endowments among countries, so that the distribution of Gini coefficients among countries may be non-uniform. Thus, we denote by $G_1, G_2, \ldots, G_n$ the samples of n different countries' Gini coefficients.

Now we assume that $G_i$ is independent of $G_j$ for any $i \neq j$. This means that there are no political or economic interventions among countries.

Denote by $\varepsilon_i = G_i - G_D$ for $i = 1, \ldots, n$ the deviation of every single country's Gini coefficient from the desirable Gini coefficient. Suppose $\varepsilon_i$'s distribution density function is $f(\varepsilon_i)$, then the joint density function for all observations is

$$L(G_D) = L(G_D; G_1, \ldots, G_n) = \prod_{i=1}^{n} f(\varepsilon_i) = \prod_{i=1}^{n} f(G_i - G_D) \qquad (D.1)$$

Set

$$\frac{d \log L(G_D)}{dG_D} = 0 \qquad (D.2)$$

Rearranging (D.2) as

$$\sum_{i=1}^{n} \frac{f'(G_i - G_D)}{f(G_i - G_D)} = 0 \qquad (D.3)$$

Substituting $g(G_i - G_D) = \frac{f'(G_i - G_D)}{f(G_i - G_D)}$ into (D.3) yields

$$\sum_{i=1}^{n} g(G_i - G_D) = 0 \qquad (D.4)$$

We assume that the solution governed by maximum likelihood estimation is exactly the arithmetic average. So we plug arithmetic average of Gini coefficient-$\bar{G}$ into (D.4) and obtain

$$\sum_{i=1}^{n} g(G_i - \bar{G}) = 0 \qquad (D.5)$$

If we set $n = 2$, then we have

$$g(G_1 - \bar{G}) + g(G_2 - \bar{G}) = 0 \qquad (D.6)$$

It is easy to see $(G_1 - \bar{G}) = -(G_2 - \bar{G})$. Since $G_1$ and $G_2$ are arbitrary, then we get

$$g(-\varepsilon) = -g(\varepsilon) \qquad (D.7)$$

We set $n = m + 1$ in (D.5) and meanwhile let $\varepsilon_1 = \cdots \varepsilon_m = -\varepsilon$ and $\varepsilon_{m+1} = m\varepsilon$ so that $\bar{\varepsilon} = 0$, then we have

$$\sum_{i=1}^{n} g(G_i - \bar{G}) = mg(-\varepsilon) + g(m\varepsilon) \qquad (D.8)$$

Substituting (D.5) and (D.7) into (D.8) yields

$$g(m\varepsilon) = mg(\varepsilon) \qquad (D.9)$$

The only continuous function $g(\cdot)$ satisfying (D.9) is $g(\varepsilon) = c\varepsilon$, so we get

$$f(\varepsilon) = Me^{c\varepsilon^2} \tag{D.10}$$

Due to $\int_{-\infty}^{+\infty} f(\varepsilon)d\varepsilon = 1$ finally we have

$$f(\varepsilon) = \frac{1}{\sqrt{2\pi\sigma^2}} e^{-\frac{\varepsilon^2}{2\sigma^2}} \tag{D.11}$$